\documentclass[twocolumn,showpacs,amsmath,amssymb]{revtex4}
\usepackage{graphicx}
\usepackage{dcolumn}
\usepackage{bm}
\usepackage{tikz}
\usepackage{hyperref}

\begin{document}
\title{Limits of heralded single photon sources based on parametric photon pair generation}

\author{St{\'e}phane Virally}
\email{stephane.virally@polymtl.ca}
\author{Suzanne Lacroix}
\author{Nicolas Godbout}
\affiliation{COPL, D{\'e}partement de G{\'e}nie Physique, {\'E}cole Polytechnique de Montr{\'e}al\\
C.P. 6079, Succursale Centre-Ville, Montr{\'e}al, Qu{\'e}bec, H3C 3A7, Canada}
\date{\today}

\begin{abstract}
We derive calculations on the statistics of a heralded single photon source based on parametric photon pair generation. These calculations highlight fundamental and practical limits for these sources, and show which physical parameters can be optimized to improve the quality of a real source.
\end{abstract}

\maketitle

\section{\label{sec:HSPS}Heralded single photon sources}
Although the concept of a light particle has existed for a long time, the modern description of what would later be termed ``photon'' was introduced by A. Einstein in 1905~\cite{Einstein02}. Since then, progress has been made in the detection of single photons~\cite{Cabrera01}, but the design of true single photon sources remains a challenge. One of the proposed solutions is a heralded single photon source (HSPS)~\cite{Fasel01, Fulconis01} based on a parametric source of photon pairs. 

We derive herein calculations taking into account relevant physical parameters of such a source. The simple HSPS setup on which these calculations are based entails no interference effects, so that we can focus on photon statistics rather than the full picture of the quantum states involved. These simple results shed light onto the importance of each individual parameter in the quality of the source, and provide formul{\ae} that can be used to find each parameter's optimum value. They also provide fundamental limits for the performance of HSPS, as these sources are inherently imperfect.

In section~\ref{sec:model}, we detail the basic HSPS model. In section~\ref{sec:statistics} we derive the exact statistics associated with this model. In section~\ref{sec:properties and limits} we examine the properties and limits of unfiltered HSPS. We then analyze the effect of filtering in section~\ref{sec:filtering}. We conclude in section~\ref{sec:conclusion}.

\section{\label{sec:model}A simple HSPS model}
\begin{figure}
\centering
\begin{tikzpicture}
\draw (0,0) rectangle (1.2,2);
\node at (0.6,1) {PPPS};
\draw [dashed] (1.4,1.5) -- (3.5,1.5);
\node at (2.45,1.8) {HB};
\draw [dashed] (1.4,0.5) -- (3.5,0.5);
\node at (2.45,0.8) {SB};
\draw [double distance=1.5pt] (4.4,1.5) -- (5.1,1.5);
\node at (4.6,1.8) {D};
\draw [dashed] (3.96,0.5) -- (5.1,0.5);
\fill (3.7,1.5) circle (0.16);
\fill (3.7,0.5) circle (0.16);
\draw (4,1.1) arc (-90:90:0.4) -- (4,1.1);
\node at (5.5,1.5) {EO};
\node at (5.5,0.5) {OO};
\end{tikzpicture}
\caption{Simplified HSPS setup. PPPS: parametric photon pair source (photons within a single pair are supposed to be distinguishable and therefore separable); HB: heralding branch; D: detector; EO: electrical output (heralding signal); SB: signal branch; OO: optical output (heralded photons).}
\label{fig:HSPS}
\end{figure}
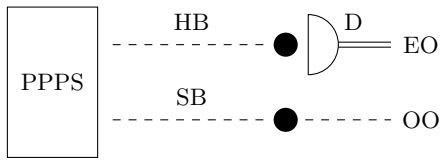

We propose to study the simple setup represented in Fig.~\ref{fig:HSPS}. In this model, a source of photon pairs (e.g., generated via parametric down conversion in a crystal~\cite{Tanzilli02} or four-wave mixing in a fiber~\cite{Fulconis01}) is assumed to statistically provide \(N\) photon pairs with probability \(P_\textrm{in}(N)\) during a specified time bin (e.g., the duration of the pump pulse or the active time of a single photon detector).

The two photons in each pair are furthermore assumed to be distinguishable and separable (e.g., using spatial or frequency filtering). One photon from each pair is used as a heralding signal for the second photon. The heralding photon is detected by a single photon detector, hence announcing the presence of another photon in the signal (heralded) branch. For a perfect HSPS, the probability of having no photon in the signal branch is reduced to zero when a heralding signal is present.

However, HSPS are fundamentally limited by physics, as even a perfect heralding system would not provide a true single photon source. Indeed, parametric processes inherently generate multiple pairs with a non-zero probability. These multiple pairs cannot be eliminated by the heralding signal. In fact, we will see below that the heralding system increases their probability. In addition, none of the physical elements of a real source are perfect. In particular, losses occur in both branches and at the detector level, and dark counts can provide false heralding signals.

Fortunately, it is rather straightforward to model losses in a quantum system. Indeed, a lossy system can always be modeled by one or several simplified beam splitters for which we consider only one input and one output each~\cite{Loudon01}. In addition, some systems (such as the one studied here) are so simple that no interference takes place. In those instances, one needs only to compute \emph{intensities} (i.e., probabilities) instead of complex \emph{amplitudes} for the quantum states. This simplification ensures a better readability of the results.

In this instance, assuming \(n_\textrm{in}=N\) photons at the input of the simplified beam splitter, \(0\leq n_\textrm{out}\leq N\) photons at the output, and a probability of transmission \(0\leq\eta\leq1\), we have~\cite{Campos01}
\begin{equation}
P(n_\textrm{out}=n)=\binom{N}{n}\eta^n(1-\eta)^{N-n},
\label{eq:beamsplitter}
\end{equation}
a simple binomial law for the transmission of \(n\) photons.

All losses in a single branch (including detection losses due to the limited efficiency of the detector) can be compounded and modeled by a single beam splitter. We assume that losses are independent of the mode (i.e., independent of the wave-vector direction, spectrum and polarization of the photons). Effects of mode-dependent filtering are analyzed separately in section~\ref{sec:filtering}.

\section{\label{sec:statistics}Statistics of the model}
Let us assume first that \(N\) pairs have been produced in a single time bin at the input of the system. Let us note \(\eta_h\) the transmission efficiency of the heralding branch (including the quantum efficiency of the detector), and \(\eta_s\) the transmission efficiency of the signal, or heralded, line. Let us also note \(d_h\) the probability of a dark count on the detector during a single time bin. Using Eq.~\ref{eq:beamsplitter}, we find the probability that a heralding signal is triggered (i.e., the detector clicks) to be
\begin{equation}
\begin{split}
H(N)&=(1-d_h)\sum_{k=1}^N\binom{N}{k}\eta_h^k(1-\eta_h)^{N-k}+d_h\\
&=1-(1-d_h)(1-\eta_h)^N.
\end{split}
\label{eq:Ph}
\end{equation}

Remembering that the probability of having \(N\) photons at the input is noted \(P_\textrm{in}(N)\), the total probability of having \(n\) photons at the output of the signal line, conditional to the presence of a heralding signal is
\begin{equation}
P_s(n)=\frac{\displaystyle\sum_{N=n}^{+\infty}\binom{N}{n}P_\textrm{in}(N)H(N)\eta_s^n(1-\eta_s)^{N-n}}{\displaystyle\sum_{N=0}^{+\infty}P_\textrm{in}(N)H(N)}.
\label{eq:Ps}
\end{equation}

We can apply Eq.~\ref{eq:Ps} to the most common type of probability for the generation of photon pairs, namely Poisson statistics. In the case of photon pairs generated via parametric down conversion or four wave mixing, this type of statistics arises when a large number of distinguishable modes are accessible to the parametric process \cite{DeRiedmatten01}. In this case, the \(N\)-photon pair probability is
\begin{equation}
P_\textrm{in}(N)=e^{-\mu}\frac{\mu^N}{N!},
\label{eq:poisson}
\end{equation}
where \(\mu\) is the average number of photons pairs generated per time bin.

Injecting Eq.~\ref{eq:Ph} and~\ref{eq:poisson} into Eq.~\ref{eq:Ps}, we get
\begin{equation}
P_s(n)=e^{-\mu\eta_s}\frac{(\mu\eta_s)^n}{n!}\xi_p(n),
\label{eq:Ps_poisson}
\end{equation}
where
\begin{equation}
\xi_p(n)=\frac{1-(1-d_h)(1-\eta_h)^ne^{-\mu\eta_h(1-\eta_s)}}{1-(1-d_h)e^{-\mu\eta_h}}
\label{eq:xip}
\end{equation}
is a correcting term to the Poisson distribution \(e^{-\mu\eta_s}(\mu\eta_s)^n/n!\) that would be observed at the output of the signal line in the absence of a heralding signal. This correcting term usually ensures that the statistics becomes sub-Poisson (as measured by a \(g^{(2)}(0)\) factor lower than 1 for the photons at the output of the heralded line).

There is also a second interesting case where Eq.~\ref{eq:Ps} can be applied. When heavy mode filtering takes place downstream of the parametric process, or in the case of some inherently narrow processes \cite{Corona01}, the statistics of the produced pairs is thermal \cite{DeRiedmatten01}. This case is very interesting, because it leads to indistinguishable photons (i.e., photons produced in the same spatial and spectral mode), which is a requirement for most linear optical quantum computing (LOQC) operations \cite{DeRiedmatten01}. The \(N\)-photon pair thermal probability is
\begin{equation}
P_\textrm{in}(N)=\frac{1}{1+\mu}\left(\frac{\mu}{1+\mu}\right)^N
\label{eq:thermal}
\end{equation}

Injecting Eq.~\ref{eq:Ph} and~\ref{eq:thermal} into Eq.~\ref{eq:Ps}, we get
\begin{equation}
P_s(n)=\frac{1}{1+\mu\eta_s}\left(\frac{\mu\eta_s}{1+\mu\eta_s}\right)^n\xi_t(n),
\label{eq:Ps_thermal}
\end{equation}
where
\begin{multline}
\xi_t(n)=\frac{1+\mu\eta_h}{d_h+\mu\eta_h}\\\left\{1-\frac{(1-d_h)(1-\eta_h)^n(1+\mu\eta_s)^{n+1}}{\left[1+\mu(\eta_s+\eta_h-\eta_s\eta_h)\right]^{n+1}}\right\}
\label{eq:xit}
\end{multline}
is a correcting factor to the thermal distribution \((\mu\eta_s)^n/(1+\mu\eta_s)^{n+1}\) that would be observed at the output of the signal line in absence of a heralding signal \footnote{Whatever the actual nature (pure or mixed) of the state at the input of the beam splitter, a state with a Poisson or thermal statistics for the arrival of photons retains that statistics after the beam splitter (even though the nature of the state can be changed).}.

\section{\label{sec:properties and limits}Properties and limits of HSPS}
\subsection{\label{sec:performances}Properties}
The form of Eqs.~\ref{eq:Ps_poisson} and~\ref{eq:xip} (resp. Eqs.~\ref{eq:Ps_thermal} and~\ref{eq:xit}) makes it straightforward to understand the significance of the physical parameters of the source.

First, we have
\begin{equation}
\frac{\xi_p(1)}{\xi_p(0)}=\frac{1-(1-d_h)(1-\eta_h)e^{-\mu\eta_h(1-\eta_s)}}{1-(1-d_h)e^{-\mu\eta_h(1-\eta_s)}}
\end{equation}
for Poisson statistics, and
\begin{multline}
\frac{\xi_t(1)}{\xi_t(0)}=\frac{1+\mu(\eta_s+\eta_h-\eta_s\eta_h)}{1+\mu(\eta_s+\eta_h-\eta_s\eta_h)^2}\\\frac{1+\mu(\eta_s+\eta_h-\eta_s\eta_h)^2-(1-d_h)(1-\eta_h)(1+\mu\eta_s)^2}{1+\mu(\eta_s+\eta_h-\eta_s\eta_h)-(1-d_h)(1+\mu\eta_s)}
\end{multline}
for thermal statistics respectively.

For the limit case \(\mu=0\) we obtain
\begin{equation}
\frac{\xi_p(1)}{\xi_p(0)}=\frac{\xi_t(1)}{\xi_t(0)}=1-\eta_h+\frac{\eta_h}{d_h}.
\end{equation}
So for \(\eta_h\gg d_h\) (i.e., when the transmission efficiency of the heralding branch is much greater than the dark count probability on the detector), there is an important increase in the probability of seeing one photon versus that of seeing no photon in the signal branch. That is the main mechanism of HSPS.

Eqs.~\ref{eq:xip} and~\ref{eq:xit} show functions \(\xi_p(n)\) and \(\xi_t(n)\) growing with \(n\). This means that multiple pair probabilities increase even more than the probability of a single pair. This is detrimental to the process of getting a true single photon source. However, this effect is small compared to the attenuation provided by a small \(\mu\eta_s\) (see Eqs.~\ref{eq:Ps_poisson} and~\ref{eq:Ps_thermal}). Indeed, \(\xi_p(n)\) and \(\xi_t(n)\) are concave and tend asymptotically towards respective finite limits
\begin{equation*}
\lim_{n\rightarrow+\infty}\xi_p(n)=\frac{1}{1-(1-d_h)e^{-\mu\eta_h}}
\end{equation*}
and
\begin{equation*}
\lim_{n\rightarrow+\infty}\xi_t(n)=\frac{1+\mu\eta_h}{d_h+\mu\eta_h}.
\end{equation*}

\begin{figure}
\centering
\includegraphics[width=0.48\textwidth]{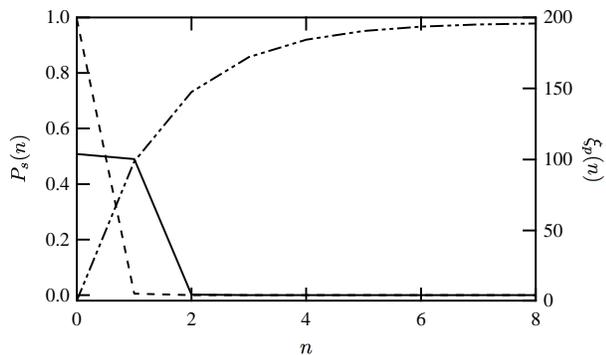}
\caption{Example of probability modifications due to heralding. On the left axis: probabilities associated with the non-heralded signal branch (dashed line); improved probabilities associated with the heralded signal (full line). On the right axis: values of \(\xi_p(n)\) (dash-dotted line). In this example, \(\eta_h=\eta_s=50\%\), \(d_h=10^{-4}\) and \(\mu=0.01\).}
\label{fig:limits}
\end{figure}

In the end, the original exponentially decreasing probability (inherent to Poisson and thermal statistics with \(\eta_s\mu\ll1\)) is multiplied by the relatively slowly growing correcting factor \(\xi\). It is then possible to increase the probability for a single photon, while keeping the chances of getting multiple pairs low. Fig.~\ref{fig:limits} shows an example of this feature in the case of Poisson statistics. In this case, the graphs for thermal statistics would be almost identical.

\subsection{\label{sec:limits}Limits}
Let us examine what happens in a perfect experiment where \(\eta_s=\eta_h=1\) and \(d_h=0\).

First, we get \(\xi_p(0)=\xi_t(0)=0\). As expected, all instances where no photons are present are eliminated.

Also, the correcting factor for a single photon in the signal branch is respectively \(\xi_p(1)=e^\mu/(e^{\mu}-1)\) and \(\xi_t(1)=(1+\mu)/\mu\). Since HSPS are usually operated in regimes where \(\mu\ll1\), these factors can be very large. However, the probability of getting exactly one photon is not unity, but respectively \(\mu/(e^{\mu}-1)\) and \(1/(1+\mu)\). These terms become unity only for \(\mu=0\), i.e., only when no photons are generated! This is a fundamental limitation of HSPS, due to the parametric process used for the generation of photon pairs.

As we have seen, the probability of getting multiple pairs is increased by the heralding process. This effect can only be managed by reducing \(\mu\). However, real sources cannot be dimmed too much because of dark counts on the detector.

In order to determine the limits of dimming, we use Eq.~\ref{eq:Ps_poisson} to compute the photon number variance of a Poisson-based HSPS as
\begin{equation}
(\Delta n)^2=\mu\eta_s\{1+\gamma\eta_h[1-\mu\eta_s\eta_h(1+\gamma)]\},
\label{eq:variance}
\end{equation}
where
\begin{equation}
\gamma=\frac{(1-d_h)e^{-\mu\eta_h}}{1-(1-d_h)e^{-\mu\eta_h}}
\label{eq:gamma}
\end{equation}

In addition, the average number of photons is
\begin{equation}
\left\langle n\right\rangle=\mu\eta_s(1+\gamma\eta_h)
\label{eq:navg}
\end{equation}

A true single-photon source would have zero variance. In practice, we want the source to exhibit strong sub-Poisson behavior, i.e. \((\Delta n)^2\ll\left\langle n\right\rangle\).

\begin{figure}
\centering
\includegraphics[width=0.48\textwidth]{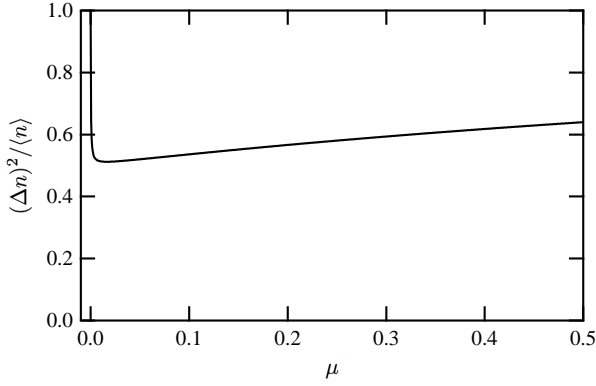}
\caption{Effect of dimming. \((\Delta n)^2/\left\langle n\right\rangle\) as a function of \(\mu\) for \(\eta_h=\eta_s=50\%\) and \(d_h=10^{-4}\). Optimal operation arises for $\mu\simeq0.016$ photon per time bin.}
\label{fig:subP}
\end{figure}

For \(\mu\rightarrow0\), dark counts on the detector become dominant and \((\Delta n)^2\rightarrow\left\langle n\right\rangle\). The same is true for \(\mu\rightarrow+\infty\), this time because the probability of having zero photons becomes negligible even in the absence of the heralding signal. Hence, minimizing \((\Delta n)^2/\left\langle n\right\rangle\) provides a value of \(\mu\) that optimizes the sub-Poisson behavior of the source, as shown on Fig.~\ref{fig:subP} for the same set of parameters as that of Fig.~\ref{fig:limits}. An equivalent curve for thermal statistics at the output of the parametric photon pair source would exhibit the same behavior, as thermal and Poisson statistics are very similar for small values of \(\mu\). 

\section{\label{sec:filtering}Effects of filtering}
A filter (e.g., spectral, spatial or polarization) can be placed in one or the other line in order to purify the heralded states (i.e., only one mode is left unfiltered and photons in the heralded branch become truly indistinguishable).

The effect of filtering is different from the effect of attenuation. Filtering one of the branches has two effects, namely selective attenuation in the filtered branch and projection of the photons in the other branch into a statistical mixture of photons paired with an unfiltered photon and extraneous photons (formerly paired with filtered-out photons).

In this section, we consider that the pair source features a Poisson statistics, and the filtering selects only one mode with a thermal statistics. The extraneous photons in the unfiltered branch then retain a Poisson statistics (i.e., the number of corresponding modes remains large). We denote by \(f\) the transmitted fraction of photons through the filter, so that \(\mu f\) is the mean number of photons allowed to go through, while \(\mu(1-f)\) is the mean number of photons filtered out of the branch.

\subsection{\label{sec:s filtering}Filtering the signal branch}
In order to purify the modes of the heralded photons, it is natural to first think about placing a filter in the signal branch. In such a configuration, the extraneous photons in the heralding branch can simply be seen as additional background noise on the detector. It is the equivalent of replacing \(d_h\) by
\begin{multline}
\nu_h=d_h\\+(1-d_h)e^{-\mu(1-f)}\sum_{k=0}^{+\infty}\frac{[\mu(1-f)]^k}{k!}\left[1-(1-\eta_h)^k\right]\\
=1-(1-d_h)e^{-\mu\eta_h(1-f)}.
\end{multline}

We then have
\begin{equation}
P_s(n)=\frac{1}{1+\mu f\eta_s}\left(\frac{\mu f\eta_s}{1+\mu f\eta_s}\right)^n\xi_s(n),
\label{eq:Ps_fs}
\end{equation}
with
\begin{multline}
\xi_s(n)=\frac{1+\mu f\eta_h}{\nu_h+\mu f\eta_h}\\\left\{1-\frac{(1-\nu_h)(1-\eta_h)^n(1+\mu f\eta_s)^{n+1}}{[1+\mu f(\eta_s+\eta_h-\eta_s\eta_h)]^{n+1}}\right\},
\label{eq:xis}
\end{multline}
as the filtered mode exhibits thermal statistics.

The filter improves the modal purity of the source. However, comparing equations~\ref{eq:Ps_thermal} and~\ref{eq:xit} with equations~\ref{eq:Ps_fs} and~\ref{eq:xis}, it can be seen that the performance of the source is affected by the presence of the filter. In addition to decreasing the number of heralded photons (a feature of filtering), the filter also increases the noise on the heralding line (as \(d_h\) is replaced by an always larger \(\nu_h\)). This additional noise increases the probability of seeing no photon in the signal branch when a heralding signal is present, which greatly reduces the gain obtained by the heralding signal in the first place.

\subsection{\label{sec:h filtering}Filtering the heralding branch}
When a filter is placed in the heralding branch, we must take into account extraneous photons in the signal branch. Hence, the probability of finding exactly \(n\) photons in the signal branch is
\begin{equation}
\sum_{k=0}^n\binom{n}{k}P_1(k)P_2(n-k),
\end{equation}
where
\begin{multline}
P_1(k)=\frac{1}{\mu f+1}\sum_{N=k}^{+\infty}\binom{N}{k}[1-(1-d_h)(1-\eta_h)^N]\\\left(\frac{\mu f}{\mu f+1}\right)^N\eta_s^{k}(1-\eta_s)^{N-k}
\end{multline}
is the probability of finding exactly \(k\) photons paired to an unfiltered heralding photon, and
\begin{multline}
P_2(k)=e^{-\mu(1-f)}\sum_{N=k}^{+\infty}\binom{N}{k}\frac{\left[\mu(1-f)\right]^N}{N!}\\\eta_s^{k}(1-\eta_s)^{N-k}
\end{multline}
is the independent probability of finding exactly \(k\) extraneous photons.

In the end, we find
\begin{equation}
P_s(n)=\frac{1}{1+\mu f\eta_s}\left(\frac{\mu f\eta_s}{1+\mu f\eta_s}\right)^n\xi_h(n),
\label{eq:Ps_fh}
\end{equation}
with
\begin{multline}
\xi_h(n)=\frac{1+\mu f\eta_h}{d_h+\mu f\eta_h}e^{-\mu\eta_s(1-f)}\\
\left\{\alpha_n-\beta_n\;\frac{(1-d_h)(1-\eta_h)^n(1+\mu f\eta_s)^{n+1}}{\left[1+\mu f(\eta_s+\eta_h-\eta_s\eta_h)\right]^{n+1}}\right\},
\label{eq:xih}
\end{multline}
\begin{equation}
\alpha_n=\mathcal{L}_n\left[-\frac{(1+\mu f\eta_s)(1-f)}{f}\right],
\end{equation}
\begin{equation}
\beta_n=\mathcal{L}_n\left\{-\frac{[1+\mu f(\eta_s+\eta_h-\eta_s\eta_h)](1-f)}{f(1-\eta_h)}\right\},
\end{equation}
and \(\mathcal{L}_n\) is the \(n^\textrm{th}\) order Laguerre polynomial defined as~\cite[Eq.~22.11.6]{Abramowitz01}
\begin{equation}
\mathcal{L}_n(x)=\frac{e^x}{n!}\frac{d^n}{dx^n}(e^{-x}x^n)=\sum_{k=0}^n\binom{n}{k}\frac{(-x)^k}{k!}.
\label{eq:Ln}
\end{equation}

As expected, Eq.~\ref{eq:xih} reduces to Eq.~\ref{eq:xit} for \(f=1\). However, for \(f<1\), \(\xi_h(n)\) no longer features a finite limit when \(n\rightarrow+\infty\) (\(\xi_h\) grows exponentially), and multi-pair probabilities increase. Indeed, as \(n\) grows, Eq.~\ref{eq:Ps_fh} becomes
\begin{multline}
P_s(n)\simeq\frac{1+\mu f\eta_h}{(1+\mu f\eta_s)(d_h+\mu f\eta_h)}\\e^{-\mu\eta_s(1-f)}\frac{\left[\mu\eta_s(1-f)\right]^n}{n!},
\label{eq:Ps_fhN}
\end{multline}
which shows that the main contribution to the multi-photon statistics is that of the extraneous photons. Fortunately, the probability of having multiple photons in the signal branch remains negligible for sufficiently small \(\mu\eta_s\).

\begin{figure}[t!]
\centering
\includegraphics[width=0.48\textwidth]{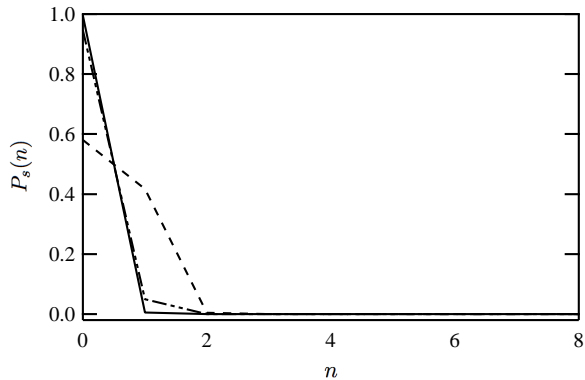}
\caption{Effects of filtering. The original, non-heralded, probability is shown as a full line. As in Fig.~\ref{fig:limits}, \(\eta_h=\eta_s=50\%\), \(d_h=10^{-4}\) and \(\mu=0.01\). A filter with \(f=10\%\) is then inserted on one of the branches. Filtering on the heralding branch is shown as a dashed line. Filtering on the signal branch is shown as a dash-dotted line.}
\label{fig:filter}
\end{figure}

Fig.~\ref{fig:filter} shows how filtering in the heralding and signal branch modifies the probabilities of the source. All filtering in a single branch is both beneficial to the purity of the modes and detrimental to the sub-Poisson quality of the source. The physical effects of filtering in the signal and heralding branches are different. Filtering in the signal branch degrades the source by increasing \(P_s(0)\), which goes against the heralding mechanism of the HSPS. On the other hand, a filter placed on the heralding branch increases multiple pair probabilities. This effect can be reduced by proper dimming of the photon pair source. Hence, filtering the heralding branch is a better solution for HSPS.
 
\section{\label{sec:conclusion}Conclusion}
We have developed a model for heralded single photon sources based on parametric photon pair generation. This model shows that the heralding process results in a multiplying factor that increases the probability of seeing at least one photon at the optical output when a heralding signal is present.

Because of the parametric photon pair source used in HSPS, the probability of seeing multiple photons at the output cannot be eliminated. In fact, the heralding process only increases that probability. Hence, the probability of having exactly one photon is never unity. The only way to reduce the occurrence of multiple pairs is to lower the average number of pairs produced by the parametric source. However, dark counts on the detector impose a lower limit on that average value.

When a modally pure source of photons is required, the best solution is to try and increase the purity of the parametric photon source. However, this is not always possible, and additional filtering might be required inside the heralding system. In this case, it is preferable to implement filtering of the heralding line, rather than the signal line.

\bibliography{arXiv}

\end{document}